\documentclass[aps,prb,reprint,floatfix,superscriptaddress]{revtex4-1}

\usepackage{amsmath,amsfonts,amssymb} %
\usepackage{mathtools} %
\usepackage{wasysym} %
\usepackage{bm} %
\usepackage[bbgreekl]{mathbbol} %
\usepackage{graphicx} %
\usepackage[dvipsnames,table]{xcolor} %
\usepackage{tabularx} %
\usepackage[acronym]{glossaries} %
\usepackage{braket} %
\usepackage{multirow} % Required for multicells

\usepackage{fancyhdr} %
\usepackage{microtype} %
\usepackage{natmove}

\usepackage[byname]{smartref} %
\usepackage[breaklinks, %
colorlinks, linkcolor=Blue, citecolor=Blue, urlcolor=Blue]{hyperref} %
\usepackage[version=3]{mhchem}
\AtBeginDocument{\usepackage{booktabs}} % Fix incompatibility with

\newcommand{\eq}[1]{Eq.~(\ref{#1})} %
\newcommand{\eqs}[1]{Eqs.~(\ref{#1})} %
\def\be{\begin{equation}} %
\def\ee{\end{equation}} %
\newcommand{\kh}{\hat \kappa}

\newcommand{\CR}[1]{\hat a^{\dagger}_{#1}}
\newcommand{\AN}[1]{\hat a_{#1}}
\newcommand{\BC}[1]{\hat \beta^{\dagger}_{#1}}
\newcommand{\BA}[1]{\hat \beta_{#1}}
\newcommand{\EU}[1]{\hat E^{#1}}
\newcommand{\ED}[1]{\hat E_{#1}}
\newcommand{\EE}[2]{\hat E_{#2}^{#1}}

\newcommand{\bea}{\begin{eqnarray}}
\newcommand{\eea}{\end{eqnarray}}
\newcommand{\LA}[1]{\mathfrak{#1}}
\newcommand{\GP}[1]{\hat \gamma_{#1}}

\newcommand{\HC}{\hat C}
\newcommand{\HA}[1]{\hat A_{#1}}
\newcommand{\BCC}[1]{{\color{black} {#1}}}

\begin{document}

%\title{Lie-algebraic approach to defining mean-field theories}
\title{How to define quantum mean-field solvable Hamiltonians using Lie algebras}

\author{Artur F. Izmaylov}
\email{artur.izmaylov@utoronto.ca}
\affiliation{Department of Physical and Environmental Sciences,
  University of Toronto Scarborough, Toronto, Ontario, M1C 1A4,
  Canada}
\affiliation{Chemical Physics Theory Group, Department of Chemistry,
  University of Toronto, Toronto, Ontario, M5S 3H6, Canada}

\author{Tzu-Ching Yen}
\affiliation{Chemical Physics Theory Group, Department of Chemistry,
  University of Toronto, Toronto, Ontario, M5S 3H6, Canada}

\begin{abstract}
Necessary and sufficient conditions for quantum Hamiltonians to be exactly solvable within mean-field theories 
have not been formulated so far. 
To resolve this problem, first, we define what mean-field theory is, independently 
of a Hamiltonian realization in a particular set of operators. Second, using a Lie-algebraic framework 
we formulate a criterion for a Hamiltonian to be mean-field solvable. The criterion is applicable for both 
distinguishable and indistinguishable particle cases.  
For the electronic Hamiltonians, our approach reveals the existence of mean-field solvable Hamiltonians of 
higher fermionic operator powers than quadratic. Some of the mean-field solvable Hamiltonians 
require different sets of quasi-particle rotations for different eigenstates, which 
reflects a more complicated structure of such Hamiltonians.  
\end{abstract}
%\date{\today}
\maketitle

\section{Introduction} 

Mean-field (MF) theories are useful for solving many-body problems in both classical and quantum computing. 
In classical computing, they provide an approximate description that can 
serve as a basis for further more accurate perturbational treatment.\cite{Helgaker} Another useful feature of MF procedures is 
the relative simplicity of their eigenstates, which can be thought of as the most general un-entangled states,\cite{Barnum:2003id} 
and thus are easy to represent on a classical computer. In quantum computing, 
formulation of MF-solvable Hamiltonians allows one to use such Hamiltonians as elementary blocks 
in decomposing realistic Hamiltonians for efficient quantum measurements.\cite{CSA2020} 
Also, it was shown that MF transformations are computationally simple and can be efficiently implemented on
both quantum and classical computers.\cite{Somma:2006ke,Somma:2019km} 

MF theories can be formulated using different sets of operators to express many-body Hamiltonians, for example, 
fermionic creation and annihilation or qubit Pauli operators, with fermion-qubit transformations 
to switch between these operator sets.\cite{Bravyi:2002/aph/210,Seeley:2012/jcp/224109} 
One may wonder what the common features of all  
MF treatments are, independent of the Hamiltonian operator expression? 
Usually, this question is addressed by introducing 
a notion of single particles and single-particle operators. Then an $N$-particle state is considered to be 
an MF-state if it is an eigenstate of a set of $N$ commuting single-particle operators. This is equivalent for the 
$N$-particle state to be a product of single-particle eigenstates of the commuting set of single-particle operators.
\footnote{Here, by a product we refer to a proper tensor product between states from different single-particle spaces. 
For example, indistinguishable boson/fermion particles requires symmetrized/anti-symmetrized products.} 
During the MF procedure, only product states are used as trial states, and thus, MF rotations are unitary operations
that transform one product state into another. Having 
a multi-particle state to be a product of single-particle states can be seen as the origin for the term {\it mean-field}.  
Indeed, each particle is described by an eigenstate of some single-particle operator as if it does not interact with 
other particles but rather is in some effective potential or a mean-field created by other particles.

However, this consideration does not provide a straightforward 
path to formulating necessary and sufficient conditions for the Hamiltonian exact solvability by the MF procedure. 
In other words, what many-body Hamiltonians can have eigenstates that are products of one-particle states?
There are well-known sufficient conditions for the MF-solvability: 
separable Hamiltonians, mainly appearing with distinguishable particles like qubits, 
and the Hamiltonians quadratic in fermionic and bosonic creation and annihilation operators.  
Yet, these cases do not cover all MF-solvable Hamiltonians. For example, it was found that for distinguishable 
particles there exist a large class of non-separable MF-solvable Hamiltonians.\cite{Izmaylov2019revising,Poirier:1997do} 

In this work, using a unifying Lie-algebraic framework we will provide necessary and sufficient conditions for the Hamiltonian 
to be MF-solvable for both distinguishable and indistinguishable particles.   
Similar Lie-algebraic consideration was done previously in Ref.~\citenum{Somma:2006ke}, but due to a 
different focus, Ref.~\citenum{Somma:2006ke} did not cover MF-solvable Hamiltonians that require multiple 
different unitary rotations to obtain all eigenstates (classes higher than 1 in our classification). 

Our definition of MF-solvable Hamiltonians is based on the condition that all eigenstates of such Hamiltonians should be MF states. 
To define general MF states, we introduce a set of operators closed with respect to commutation, the Lie algebra, whose elements are sufficient 
for expressing the Hamiltonian. These Lie algebra operators constitute mathematical generalization of single-particle operators.
The Lie algebra contains subsets of fully commuting operators, it is convenient to use one of the largest 
fully-commuting subsets or a Cartan sub-algebra (CSA). The MF states are defined as all states that can be constructed from eigenstates of a 
CSA by unitary transformations obtained using exponentiation of Lie algebra operators. Such MF states cover Slater determinants in 
the Hartree-Fock method\cite{Helgaker} and product states in qubit mean-field\cite{Ryabinkin:2018di} and 
Hartree-Fock-Bogoliubov\cite{Book/Ring:1980} theories. These states can be also defined for an 
arbitrary Lie algebra that is chosen for expression of the Hamiltonian. 

The main applicational value in the general formulation of the MF-solvability originates from advantages  
in changing one set of operators (forming a Lie algebra) to another in many-body Hamiltonians. It was found 
that MF theories can perform differently depending on the operator realizations for the same Hamiltonian.\cite{Ryabinkin:2018di} 
Moreover, there are examples of Hamiltonians that are not MF-solvable in one realization but are MF-solvable in another operator 
realization (e.g. $XY$-model\cite{Lieb:1961xy}). Also, with the progress in quantum hardware, it becomes necessary to reformulate various 
many-body Hamiltonians in terms of qubit operators\cite{Seeley:2012/jcp/224109,Jordan:1928/zphys/631,Tranter:2015/ijqc/1431} and to be able to transfer many-body methods between different realizations.

The rest of the paper is organized as follows. Section II presents Lie-algebraic framework for quantum Hamiltonians, defines MF-solvable 
Hamiltonians of different classes, and comments on computational procedures identifying the MF-solvable Hamiltonians 
within the Hamiltonian of interest. Section III contains applications of the MF formalism for fermionic and qubit realizations of the 
electronic molecular Hamiltonian. Section IV concludes and provides further outlook.  

\section{Theory} \label{sec:T}

\subsection{Lie-algebraic framework}

Any Hamiltonian can be written as a polynomial expression of some elementary linear operators $\{\HA{k}\}$
\bea \label{eq:H}
\hat H = \sum_{k} c_k \hat A_k + \sum_{kk'} d_{kk'} \hat A_k \hat A_{k'} +...,
\eea

\noindent where $c_k$ and $d_{kk'}$ are some constants. Choosing $\{\HA{k}\}$ appropriately allows one to have 
a set that is closed with respect to the commutation operation
\bea\label{eq:acl}
[\hat A_i,\hat A_j] = \sum_k \xi_{ij}^{(k)} \hat A_k, 
\eea
where $\xi_{ij}^{(k)} $ are so-called structural constants from the number field $\mathbb{K}$, thus, 
$\{\HA{k}\}$ forms a Lie algebra, $\mathcal{A}$.\cite{Gilmore:2008} 
Note that products like $\hat A_k \hat A_{k'}$ and higher powers of $\mathcal{A}$ elements do not generally belong 
to the Lie algebra $\mathcal{A}$, instead they are part of a universal enveloping algebra (UEA), $\mathcal{E_A}$, 
which is built as a direct sum of tensor powers of the Lie algebra 
\bea
\mathcal{E_A} = \mathbb{K}\oplus\mathcal{A}\oplus(\mathcal{A}\otimes\mathcal{A})\oplus ...,
\eea
where the Lie bracket operation is equivalent to the commutator. 
Thus, any Hamiltonian is an element of some UEA. 

Among various Lie algebras that can be chosen for realization of $\hat H$, 
for our purpose, it will be convenient to use reductive Lie algebras as $\mathcal{A}$. 
There are two reasons for this choice. First, a reductive algebra is a direct sum of abelian 
and semisimple Lie algebras, and structural theories for both of these algebra types are well-understood.\cite{Barut:1980} 
Second, in Appendix A we show that field $\mathbb{K}$ can be chosen to be real for $\mathcal{A}$ 
realizing $\hat H$, and thus, $\mathcal{A}$ is a compact Lie algebra, which allows us to use 
some powerful theorems for this type of algebras. Since the compact Lie algebras are always reductive, 
Appendix A also shows that it is possible to realize the Hamiltonian using reductive Lie algebras.    

Any reductive Lie algebra $\mathcal{A}$ has a maximal abelian sub-algebra $\mathcal{C}\subset \mathcal{A}$ 
that is referred to as the CSA. 
If $\mathcal{A}$ is a direct sum of abelian $\LA{a}$ and semisimple $\LA{s}$ algebras, then 
 $\mathcal{C}$ will be a direct sum of $\LA{a}$ and a CSA of $\LA{s}$. The latter is a maximal 
abelian sub-algebra of $\LA{s}$ whose elements are ad-diagonalizable.\cite{Barut:1980} 
We will denote elements of $\mathcal{C}$ as $\hat C_k$'s.  
The UEA constructed from $\mathcal{C}$, $\mathcal{E_C}$, is abelian as well.
Thus, in principle, all elements of $\mathcal{E_C}$ have a common set of eigenfunctions, $\ket{C_1,...C_{N}}$:
$\HC_k \ket{C_1,...C_{N}} = C_k \ket{C_1,...C_{N}}$.  
These eigenfunctions can be used to construct functional spaces for representation of $\mathcal{A}$ and operators 
from $\mathcal{E_A}$.  

In all Lie algebras that are used for expressing many-body problems, common eigenstates of $\mathcal{C}$ are 
relatively easy to obtain. Therefore, solving the eigenvalue problem for the Hamiltonian in \eq{eq:H} can be done by 
 finding a unitary operator $\hat U$ that transforms $\hat H$ into a polynomial over the CSA elements 
 \bea \label{eq:UHU}
\hat U\hat H \hat U^\dagger= \sum_{k} c_k \hat C_k + \sum_{kk'} d_{kk'} \hat C_k \hat C_{k'} +...
 \eea
Hamiltonians are hermitian operators, which makes them diagonalizable. Therefore, $\hat U$ always exist
since \eq{eq:UHU} provides such a diagonal form of the Hamiltonian in the basis of the CSA eigen-states. 
The difficulty is that for a general Hamiltonian, $\hat U$ is an element of the Lie group corresponding 
to the exponential map of the exponentially large Lie algebra obtained by commutator closure of the UEA. Thus, it can 
be very difficult to find such $\hat U$ in general.   

\subsection{Mean-field solvable Hamiltonians}

To define a set of the MF solvable Hamiltonians we introduce a restricted class of unitary transformations, MF rotations
\bea\label{eq:Us}
\hat U_{\rm MF} = \prod_{k=1}^{|\mathcal{A}|} e^{\theta_k\hat A_k},
\eea
where $\theta_k$  are parameters defining MF rotational angles, 
$\hat A_k$ are anti-hermitian operators, and $|\mathcal{A}|$ is the number of $\hat A_k$ generators in 
$\mathcal{A}$.  $\hat U_{\rm MF}$'s form the universal covering 
Lie group corresponding to the Lie algebra $\mathcal{A}$. In spite of general 
non-commutativity of generators $\hat A_k$, due to the algebraic closure according to \eq{eq:acl}, any pair of $\hat U_{\rm MF}$'s that are different by 
the order of exponents in \eq{eq:Us} can be made equal by selecting $\theta_k$'s in one of these $\hat U_{\rm MF}$'s.\cite{Izmaylov:2020hb}
\BCC{Another property of the MF unitaries that stems from the algebraic closure is 
\bea
\hat U_{\rm MF}^\dagger \hat A_{k'} \hat U_{\rm MF} = \sum_k c_k \hat A_k,
\eea
where $\hat A_{k'}$ is any Lie algebra element.
This property is essential for computational advantage of MF theories since it 
results in preservation of the degree for any polynomial function of algebra elements $\hat A_k$ upon
transformation by $\hat U_{\rm MF}$. }

%To determine whether we assume that the Hamiltonian of interest is realized using the Lie algebra $\mathcal{A}$. 
We define the mean-field solvable Hamiltonians as the ones that have the form 
\bea \label{eq:GMF}
\hat H_{\rm MF} = \sum_{J}  \hat V_J \ket{\bar{C}_J}E_J\bra{\bar{C}_J} \hat V_J^\dagger,
\eea
where $E_J$ are eigen-values, $\ket{\bar{C}_J} = \ket{C_1^{(J)},...C_N^{(J)}}$ are basis states that are eigen-states of the CSA operators, 
and $\hat V_J$ are unitary MF rotations (\eq{eq:Us}). Here, any basis state $\ket{\bar{C}_J}$ can be transformed 
to the eigenstate of $\hat H_{\rm MF}$ using the MF rotation $\hat U_{\rm MF} = \hat V_J$. 
It is convenient to separate \eq{eq:GMF} Hamiltonians into classes, class $K$ Hamiltonians 
contain $K$ different unitary transformations $\{\hat V_J\}_{J=1}^K$. 
%The class number corresponds to the number of different $\hat V_J$'s, for example, class 1 corresponds to a single $\hat U = \hat V_J$ 
%that rotates all $\ket{\bar{C}_J}$ into eigenstates of $\hat H_{\rm MF}$. 
For all classes, due to hermiticity of $\hat H_{\rm MF}$, 
there is the orthogonality condition
\bea\label{eq:Or}
\bra{\bar{C}_J}\hat V_J^\dagger\hat V_I \ket{\bar{C}_I} = \delta_{JI}. 
\eea

The form given by \eq{eq:GMF} is sufficient for the MF-solvability 
because all its eigenstates can be obtained using the MF rotations. %, $\hat V_J \ket{\bar{C}_J}$.  
It is also a necessary condition because if a hermitian Hamiltonian has eigenstates in the MF form $\hat V_J \ket{\bar{C}_J}$, 
then it can be always written as \eq{eq:GMF} with the condition that all eigenstates are orthogonal. 
The main shortcoming of \eq{eq:GMF} as a criterion for determining the MF-solvability of an arbitrary Hamiltonian 
is that it does not employ only the elements of the Lie algebra but also requires projections $\ket{\bar{C}_J}\bra{\bar{C}_J}$ 
and the orthogonality condition (\eq{eq:Or}). 
In what follows we will express projectors as functions of CSA elements and introduce the necessary orthogonality 
with minimal constraints on the MF unitaries.  
Fully algebraic definitions of MF-solvable Hamiltonians will be provided for each class separately. 

%those in we require their $\hat U$'s to be an element of the universal covering 
%Lie group obtained by exponentiation of the Lie algebra $\mathcal{A}$

{\it Class 1:} This class corresponds to a single unitary transformation $\hat V_J= \hat U_{\rm MF}^\dagger$, hence \eq{eq:GMF} can be written as 
\bea\label{eq:1}
\hat H_{{\rm MF},1} = \hat U_{\rm MF}^\dagger \sum_J \ket{\bar{C}_J}E_J\bra{\bar{C}_J} \hat U_{\rm MF}.~
\eea
The sum over $J$ can be seen as an arbitrary operator in the eigen-subspace of the CSA, therefore it can be also
written as a general Taylor series over the CSA elements
 \bea
\sum_J \ket{\bar{C}_J}E_J\bra{\bar{C}_J} &=& \sum_{k} c_k \hat C_k + \sum_{kk'} d_{kk'} \hat C_k \hat C_{k'} +...~~\\
&=& F(\hat C_k).
 \eea
 Thus, by substituting the $J$-sum in \eq{eq:1} by the Taylor expansion, we obtain
 \bea\label{eq:MF1}
\hat H_{\rm MF,1} = \hat U_{\rm MF}^\dagger F(\HC_k) \hat U_{\rm MF},
\eea
hence, \eq{eq:MF1} is the algebraic form equivalent to \eq{eq:GMF} for class 1. 

{\it Class 2:} According to \eq{eq:GMF}, for class 2, there are two MF unitary transformations $\hat V_1$ and $\hat V_2$ that create 
two sets of eigen-states by acting on two groups of basis states, $\{\ket{\bar{C}_{J_1}}\}$ and $\{\ket{\bar{C}_{J_2}}\}$.
 Let us introduce projectors on subspaces $\{\ket{\bar{C}_{J_1}}\}$ and $\{\ket{\bar{C}_{J_2}}\}$
\bea
\hat P_1 &=& \sum_{J_1} \ket{\bar{C}_{J_1}}\bra{\bar{C}_{J_1}}
\eea
 and $\hat P_1^{\perp} = 1 - \hat P_1$. 
 % on the subspace spanned by $\{\ket{\bar{C}_{J_1}}\}$ 
% and its orthogonal complement ($\{\ket{\bar{C}_{J_2}}\}$).
Individual projectors $\ket{\bar{C}_{J_1}}\bra{\bar{C}_{J_1}}$ can be written in the algebraic form using L\"owdin's projection formula
\bea
\ket{\bar{C}_{J_1}}\bra{\bar{C}_{J_1}} &=& \prod_k \prod_{C_k\ne C_k^{(J_1)}}\frac{\HC_k-C_k}{C_k^{(J_1)}-C_k}. 
\eea

% definitions
The introduced projectors allow us to formulate the algebraic form of the class-2 MF-solvable Hamiltonians 
\bea\label{eq:HMF2g}
\hat H_{\rm MF,2} = \hat U_1^\dagger \left(F_1(\HC_k)\hat P_1 +  U_2^\dagger F_2(\HC_k)\hat U_2 \hat P_1^{\perp}\right)\hat U_1,\quad
\eea
where $\{\hat U_i\}_{i=1,2}$ are MF unitaries and $\{ F_i (\HC_k)\}_{i=1,2}$ are analytic functions of CSA elements.
Here, based on $\HC_k$ eigenvalues after the $\hat U_1$ transformation, all states are partitioned in two subspaces
$\{\ket{\bar{C}_{J_1}}\}$ and $\{\ket{\bar{C}_{J_2}}\}$. The states from the first subspace acquire their eigenvalues according 
to the $F_1 (\HC_k)$ function, and the states from the second subspace rotated with $\hat U_2$ so that the eigenvalues 
of the rotated states would be determined by $F_2 (\HC_k)$. 

Hamiltonian $\hat H_{\rm MF,2}$ is hermitian because $F_1(\HC_k)$ and $\hat U_2^\dagger F_2(\HC_k)\hat U_2$
commute with $\hat P_1$ and $\hat P_1^{\perp}$. The commutation of $\hat P_1^{\perp}$ with 
$\hat U_2^\dagger F_2(\HC_k)\hat U_2$ is not satisfied by a general $\hat U_2$ transformation, and thus 
it introduces a constraint on the $\hat U_2$ transformation. To satisfy this constraint it is sufficient to remove $\hat A_k$'s in $\hat U_2$ 
that do not commute with $\hat P_1^{\perp}$.

% necessary condition
To prove that \eq{eq:HMF2g} can be a criterion for the class-2 MF-solvability, we show that it is equivalent to \eq{eq:GMF} for class 2.
Two unitaries of \eq{eq:HMF2g}, $\hat U_1$ and $\hat U_2$, should be used as
$\hat U_1^\dagger$ and $(\hat U_2\hat U_1)^\dagger$ to obtain all eigenstates. Thus, we associate 
$\hat V_1$ with $\hat U_1^\dagger$ and $\hat V_2$ with $(\hat U_2\hat U_1)^\dagger$, 
then $\hat U_2 = \hat V_2^\dagger \hat V_1$. The commutativity of $\hat U_2$ 
with the $\hat P_1$ projector is necessary to maintain the orthogonality between the $\{\ket{\bar{C}_{J_1}}\}$ and $\{\ket{\bar{C}_{J_2}}\}$ subspaces
\bea\notag
\bra{\bar{C}_{J_1}}\hat V_1^\dagger \hat V_2\ket{\bar{C}_{J_2}} &=& \bra{\bar{C}_{J_1}}\hat U_1 \hat U_1^\dagger\hat U_2^\dagger\ket{\bar{C}_{J_2}} \\ \notag
&=& \bra{\bar{C}_{J_1}}\hat U_2^\dagger\ket{\bar{C}_{J_2}} = \bra{\bar{C}_{J_1}}\bar{C}_{J_2}\rangle = 0.
\eea  
Indeed, orthogonal subspaces $\{\ket{\bar{C}_{J_1}}\}$ and $\{\ket{\bar{C}_{J_2}}\}$ correspond to 
the $\hat P_1$ and $\hat P_1^\perp$ projectors, hence, commutativity $[\hat U_2,\hat P_1]=0$ is necessary 
and sufficient for $\hat U_2^\dagger$ to have $\{\ket{\bar{C}_{J_2}}\}$ as an invariant subspace, 
and thus not to disrupt the orthogonality. 

%=====================
{\it Higher classes:} Further generalizations to higher classes can be done recursively. For class 3, the subspace 
$\{\ket{\bar{C}_{J_2}}\}$ needs to be partitioned further onto two subspaces with corresponding 
projectors $\hat P_2$ and $\hat P_2^{\perp}$. The algebraic form of class-3 MF-solvable Hamiltonian is obtained from  
\eq{eq:HMF2g} using the substitution
\bea\label{eq:HMF3g} 
F_2(\HC_k)\rightarrow F_2(\HC_k)\hat P_2 + \hat U_3^\dagger F_3(\HC_k)\hat U_3 \hat P_2^{\perp},\quad 
\eea
where $\hat U_3$ is a new MF transformation commuting with $\hat P_2^{\perp}$, $F_3$ is an analytic function of CSA elements, 
and $\hat P_2^{\perp}$ projects out the states that can be obtained by the $(\hat U_2\hat U_1)^\dagger$ transformation. 

Proving the equivalence of algebraic forms of the MF-solvable Hamiltonians of higher classes (e.g. \eq{eq:HMF3g}) to 
\eq{eq:GMF} can be done by straightforward extension of the arguments given for class 2. Thus, as the 
criterion for the MF-solvability, one can use either \eq{eq:GMF} with the extra orthogonality condition in \eq{eq:Or} 
or algebraic expressions (e.g. \eqs{eq:MF1}, \eqref{eq:HMF2g}) whose form depends on the class and where the orthogonality  
condition is substituted by the commutativity requirement between projectors and unitary transformations.  

%provide sufficient conditions for 
%MF-solvable Hamiltonians of class 2 and higher. We have also shown that these equations are also necessary conditions 
%for the Hamiltonian to be MF-solvable, and thus they provide the MF-solvability criterion for each class.   

\subsection{Computational aspect}

Using a particular Lie algebra it is easy to construct an MF-solvable Hamiltonian of any class. The problem of identifying 
whether a particular Hamiltonian is MF-solvable, and what its class is, can be efficiently addressed by finding 
its eigenstates using the variational approach. The only caveat is that it is necessary to find all MF solutions of a given 
Hamiltonian, since it is possible to envision the Hamiltonian that has only a few MF solutions and non-MF states for the rest of 
the spectrum. Appendix B presents an example of the Hamiltonian that is only partially solvable by the MF procedure
and discusses a procedure for establishing partial or full MF-solvability. If the Hamiltonian is realized using only lower powers 
of the corresponding UEA (e.g., two-electron Hamiltonians) then the algorithm for establishing MF-solvability and finding all MF 
solutions requires a non-linear optimization with a polynomial number of parameters with the number of particles. This makes 
low-degree MF-solvable Hamiltonians classically tractable. 

\BCC{In the case when it is known that the Hamiltonian is MF-solvable,
finding the lowest eigenstate and its energy is equivalent to minimization of the Hamiltonian energy functional with respect to 
$\theta_k$ parameters in \eq{eq:Us}. For a non-degenerate ground state, the MF solvability guarantees that the energy function has 
only one global minimum corresponding to the ground state. All other states correspond to saddle points in the energy function, and thus,
common optimization procedures will usually converge to the ground state in this case.} 

Another practically interesting question is related to decomposition of a given Hamiltonian in terms of MF-solvable Hamiltonians.
This problem was addressed in Ref.~\citenum{CSA2020} for class 1, where it was shown that if the Hamiltonian contains only a few 
lower powers of $\mathcal{A}$ elements in \eq{eq:H}, the decomposition procedure can be done in polynomial efforts with respect to 
the size of $\mathcal{A}$. The decomposition with higher classes has not been addressed yet, and it appears to be an exponentially 
hard problem in general because the projectors in \eqs{eq:HMF2g} and \eqref{eq:HMF3g} can contain high powers of CSA 
elements. Polynomial scaling of such decompositions is only possible if the overall power of $\hat A_k$ elements in \eq{eq:H} is restricted. 
    
\section{Applications}\label{sec:app}

Here we will illustrate how the Lie-algebraic framework for identifying the MF-solvable Hamiltonians is applicable to a 
few commonly used realizations for the electronic Hamiltonian. Historically, all these realizations had their MF 
procedures developed from the quasi-particle (Fermi liquid) types of considerations. These considerations can be related to Lie algebras,
but the relation is not obvious and will be further clarified.  

Besides illustration of MF-solvable Hamiltonians, we will use well-developed structural theory for reductive Lie algebras to produce
standard ladder operators, which provide a simple way to generate excited states starting from the ground state. 
For any reductive Lie algebra, there are only two types of operators, the ones that form the CSA $\{\hat C_k\}$
and the ladder operators $\{\hat L_j^{\pm}\}$
\bea\label{eq:Lop}
~[\hat C_k, \hat L_j^{\pm}] = \pm \alpha_{jk} \hat L_j^{\pm},
\eea 
where $\alpha_{jk}$ are constants. $\{\hat L_j^{\pm}\}$ allow one to generate the entire set of eigenstates starting from a 
maximum (minimum) weight state by using lowering (raising) ladder operators. 
The only necessary condition for generating the ladder operators is to consider the complex extension of the original compact Lie algebra,
which is achieved by extending $\mathbb{K}$ to $\mathbb{C}$.

\subsection{Fermionic algebras}

The electronic Hamiltonian $\hat H_e$, starting from its second quantized form 
can be realized using the elements of different Lie algebras.
For example, the common second quantized form of $\hat H_e$ is  
\bea \label{eq:H1}
\hat H_e = \sum_{pq} h_{pq} \CR{p}\AN{q} + \sum_{pqrs} g_{pq,rs} \CR{p}\CR{q}\AN{r}\AN{s}
\eea
where $h_{pq}$ and $g_{pq,rs}$ are real constants. 
A set of $2N$ $\CR{p},\AN{q}$ operators, where $N$ is the number of spin-orbitals, is not closed with respect to commutation
\bea
~[\CR{p},\CR{q}] &=& 2 \CR{p}\CR{q}\\
~[\AN{p},\AN{q}] &=& 2 \AN{p}\AN{q} \\ 
~[\CR{p},\AN{q}] &=& 2 \CR{p}\AN{q} - \delta_{pq}.
\eea
To transform this set to a Lie algebra one needs to add all possible products 
$\EU{pq} = \CR{p}\CR{q}$, $\ED{pq}= \AN{p}\AN{q}$, and $\EE{p}{q}= \CR{p}\AN{q}$. This addition 
provides the following commutation relation illustrating the algebraic closure
\bea
~[\EE{p}{q},\EE{r}{s}] &=& \EE{p}{s}\delta_{qr} - \EE{r}{q}\delta_{ps} \\
~[\EE{p}{q},\ED{rs}] &=& \ED{qr}\delta_{ps} - \ED{qs}\delta_{pr} \\
~[\EU{pq},\ED{rs}] &=&\EE{q}{r}\delta_{ps} + \EE{p}{s}\delta_{qr}- \EE{q}{s}\delta_{pr} - \EE{p}{r}\delta_{qs} \\
~[\ED{pq},\ED{rs}] &=& [\EU{pq},\EU{rs}]  = 0 
\eea
and 
\bea
~[\AN{p},\EE{r}{s}] &=& \delta_{pr} \AN{s} \\
~[\AN{p},\ED{rs}] &=& [\CR{p},\EU{rs}] = 0  \\
~[\AN{p},\EU{rs}] &=& \delta_{pr}\CR{s} - \delta_{ps}\CR{r}. 
\eea  
The operators $\{ \CR{p},\AN{p},\EE{p}{q}, \EU{pq}, \ED{pq}\}$ 
are not standard generators of the compact $\LA{so}(2N+1)$ Lie algebra, 
but their linear combinations are sufficient for constructing these generators.\cite{Fukutome:1981/65,PaldusSarma1985}
This construction is facilitated by introducing the Majorana operators
\bea
\GP{2p-1} &=& i(\AN{p} - \CR{p})/\sqrt{2} \\
\GP{2p} &=& (\CR{p} + \AN{p})/\sqrt{2},
\eea
which allow us to introduce the $\LA{so}(2N+1)$ generators as
\bea\label{eq:CL1}
\hat S_{j0} &=& -\hat S_{0j} = -i\gamma_j, ~ (j=1,...2N)\\ \label{eq:CL2}
\hat S_{jk} &=& [\gamma_j,\gamma_k]/4 = \gamma_j\gamma_k/2 ~ (j\ne k; j,k=1,...2N).
\eea
Generators $\hat S_{jk}$ ($j,k=0,1...,2N$) satisfy regular $\LA{so}(2N+1)$ commutation relations
\bea
[\hat S_{ij},\hat S_{kl}] &=& \delta_{jk}\hat S_{il} + \delta_{il} \hat S_{jk} - \delta_{ik} \hat S_{jl} - \delta_{jl}\hat S_{ik}.
\eea
There are $2N(2N+1)/2 = 2N^2+N$ of $\hat S_{jk}$ operators in total. 

The $\LA{so}(2N+1)$ algebra contains smaller sub-algebras that can be used to realize the Hamiltonian: 
$\LA{so}(2N)$ and $\LA{u}(N)$.\cite{Fukutome:1981/65} 
$\LA{so}(2N)$ requires only $\{\EE{p}{q}, \EU{pq}, \ED{pq}\}$ operators which can form $\hat S_{jk}$ ($j,k=1...,2N$) generators and give rise to 
the following realization of the Hamiltonian 
\bea
\hat H_{e,2} = \sum_{pq} h_{pq} \EE{p}{q} + \sum_{pqrs} g_{pq,rs} \EU{pq}\ED{rs}.
\eea

Lie algebra $\LA{u}(N)$ realization is obtained using the following transformations:
\bea\notag
\hat H_{e,3} &=& \sum_{pq} h_{pq} \CR{p}\AN{q} + \sum_{pqrs} g_{pq,rs} (\CR{p}\AN{s}\delta_{rq} - \CR{p}\AN{r}\CR{q}\AN{s}) \\
&=& \sum_{pq} [h_{pq}+\sum_r g_{pr,rq}] \CR{p}\AN{q} \notag \\ \notag
&-& \sum_{pqrs} g_{pq,rs} \CR{p}\AN{r}\CR{q}\AN{s} \\
&=& \sum_{pq} \tilde{h}_{pq} \EE{p}{q} - \sum_{pqrs} g_{pq,rs} \EE{p}{r}\EE{q}{s}. 
\eea
$\{\EE{p}{q}\}$ operators can be transformed into the $\LA{u}(N)$ compact generators  
\bea
\kh_{pq} &=& (\EE{p}{q} - \EE{q}{p})/2 \\
\kh_{pq}' &=& i(\EE{p}{q} + \EE{q}{p})/2.
\eea

%%%%%%%%%%%%%%%%%%%%%%%%%
\subsubsection{$\LA{u}(N)$ algebra}

Lie algebra $\LA{u}(N)$ can be decomposed into direct sum of $\LA{u}(1)$ and $\LA{su}(N)$. 
A $\LA{u}(1)$ generator commutes with all elements of $\LA{u}(N)$ and can be expressed as $i\hat N = i\sum_p \EE{p}{p}$. 
%The $\LA{su}(N)$ algebra contains $N^2-1$ elements $\kh_{pq}$, $\kh_{pq}'$ ($p\ne q$, $N^2-N$ elements), 
%and $i[\EE{p}{p} - \hat N / N]$ ($N-1$ elements).
The CSA for $\LA{u}(N)$ is $N$ elements $i\EE{p}{p}$, which are anti-hermitian versions of the usual 
spin-orbital occupation operators. The ladder operators can be defined as $\hat L_q^{+} = \EE{q}{p}$  and
 $\hat L_q^{-} = \EE{p}{q}$ with $\alpha_{qp}=1$ in \eq{eq:Lop}. Note that for the complex extension of $\LA{u(n)}$ 
 one can take $\hat C_p = \EE{p}{p}$ to avoid extra imaginary units in definitions. 

The components of MF-solvable Hamiltonians in this algebra are one-particle unitary transformations 
\bea\label{eq:fmf}
\hat U = \prod_{p\ne q} e^{\hat \kappa_{pq} \theta_{pq}}e^{\hat \kappa_{pq}' \phi_{pq}}
\eea
and the CSA elements $\HC_p = i\EE{p}{p}$. Appendix C presents two nontrivial (non-one-electron) $\LA{u}(N)$ 
MF-solvable Hamiltonians of class 1 and 2.

\subsubsection{ $\LA{so}(2N)$ and $\LA{so}(2N+1)$ algebras}
\label{sec:so}

The CSA for $\LA{so}(2N)$ and $\LA{so}(2N+1)$ is 
the same and contains any $N$ elements with non-overlapping indices, for example 
$\mathcal{C_S} = \{\hat S_{12}, \hat S_{34}, .... \hat S_{(2N-1)2N}\}$. Addition of any $\hat S_{0j}$ element is not feasible 
due to unavoidable index overlap. $\mathcal{C_S}$ can be connected with the $\LA{u}(N)$ CSA by 
expressing the elements in $\CR{p}$ and $\AN{q}$ operators:
\bea\notag
\hat S_{(2p-1)2p} &=& \frac{1}{2}\GP{2p-1}\GP{2p} = \frac{i}{4}(\AN{p}-\CR{p})(\CR{p}+\AN{p}) \\
&=& \frac{i}{4}\left(1-2\EE{p}{p}\right).
\eea

For the complex extension of the $\LA{so}$ algebras, it is possible to use $\hat L_p^{\pm}$ obtained for $\LA{u}(N)$, 
but there are also additional ladder operators with $\ED{pq}$ ($\EU{qp}$) 
\bea
~[\EE{p}{p},\ED{pq}] = -\ED{pq}, &\quad& [\EE{q}{q},\ED{pq}] = -\ED{pq}\\
~[\EE{p}{p},\EU{qp}] = \EU{qp},  &\quad& [\EE{p}{p},\EU{qp}] = \EU{qp}
\eea
and $\AN{p}$ ($\CR{p}$) for $\LA{so}(2N+1)$ only
\bea
~[\EE{p}{p},\AN{p}] &=& -\AN{p} \\
~[\EE{p}{p},\CR{p}] &=& \CR{p}.  
\eea

Another popular set of CSAs that can be built here, are those that do not conserve the number of particles. 
Using the Hartree-Fock-Bogoliubov\cite{Book/Ring:1980}
unitary transformation conserving the anti-commutation relations of $\CR{p}$ and $\AN{q}$ operators, 
one can introduce linear combinations
\bea
\BC{q} &=& \sum_p U_{pq} \CR{p} + V_{pq} \AN{p} \\
\BA{q} &=& \sum_p  V_{pq}^{*} \CR{p} + U_{pq}^{*}\AN{p},
\eea
where $U$ and $V$ matrices satisfying the following conditions:
\bea
U^\dagger U + V^\dagger V &=& 1\\
U U^\dagger + V^*V^T &=& 1\\
U^T V + V^T U &=& 0\\
U V^\dagger + V^* U^T &=& 0.
\eea
A set of $\{i\BC{p}\BA{p}\}$ forms a CSA for the $\LA{so}(2N)$ and $\LA{so}(2N+1)$ algebras, 
while quadratic and linear operators constructed out of $\BC{p}$ and $\BA{q}$ form 
the ladder operators analogously to the same construction with $\CR{p}$ and $\AN{q}$ operators.

\subsubsection{General connection between Clifford and Lie algebras via Majorana operators}

Majorana operators are convenient because they easily connect quasi-particle-like pictures  
and Lie-algebraic frameworks. Any set of quasi-particle fermionic operators $\{\CR{p}\}$ or $\{\AN{p}\}$ form 
Grassmann algebras: $\{\CR{p},\CR{q}\}=0$ and $\{\AN{p},\AN{q}\}=0$, but combinations of these sets do not,
since $\{\AN{p},\CR{q}\}=\delta_{pq}$.  
On the other hand, Majorana operators make proper combinations of these sets to form a Clifford algebra:
$\{\GP{p},\GP{q}\} = \delta_{pq}$. Clifford algebras with $M$ generators always produce $\LA{so}(M+1)$
Lie algebras from the commutator closure, this process was illustrated in \eqs{eq:CL1} and \eqref{eq:CL2}.  
The CSA of the resulting $\LA{so}(M+1)$ algebra can always be selected in the part which is quadratic in 
Majorana operators, and this part forms the $\LA{so}(M)$ algebra. One can always organize the quasi-particle operators (or ladder operators)
using a linear combinations of Majorana operators that return back to $\CR{p}$- and $\AN{q}$-like operators. 
These operators can be rotated by the quadratic ($\LA{so}(M)$) elements 
\bea
e^{\theta\GP{q}\GP{r}} \GP{p} e^{-\theta\GP{q}\GP{r}} = \sum_s c_s(\theta)\GP{s}
\eea
into some linear combinations of linear Majorana terms. This is based on the commutation relation 
$[\GP{p},\GP{q}\GP{r}] = 2(\GP{r}\delta_{pq} - \GP{q}\delta_{pr})$, which guarantees production of linear terms. 
 Thus, mean-field type of theories can be easily built if one has $M$ Clifford algebra elements, and it will guarantee existence of 
 the $\LA{so}(M+1)$ Lie algebra with $\lfloor M/2 \rfloor$ CSA elements. Similar analysis was done in Ref.~\citenum{Chapman}
 using graph theory techniques and qubit representation for the Hamiltonian.    

\subsection{Qubit algebras}

%%%%%%%%%%%%%%%%%%%%

Another class of realizations of the electronic Hamiltonian 
can be obtained by mapping fermionic operators to qubits using 
the Jordan-Wigner, Bravyi-Kitaev, or similar fermionic-qubit mappings.
\cite{Bravyi:2002/aph/210, Seeley:2012/jcp/224109,Tranter:2015/ijqc/1431,Setia:2017/ArXiv/1712.00446,Havlicek:2017/pra/032332} Here, we will use the 
Jordan-Wigner mapping as the simplest for illustrative purpose
\bea
\AN{p} &=& (\hat x_p-i\hat y_p)\otimes \hat z_{p-1} \otimes \hat z_{p-2} ...  \otimes \hat z_{1} \\
\CR{p} &=& (\hat x_p+i\hat y_p)\otimes \hat z_{p-1} \otimes \hat z_{p-2} ...  \otimes \hat z_{1}.
\eea
This mapping produces 
\bea\label{eq:Hq}
\hat H_q &=& \sum_{k} c_k \hat P_k, \\
\hat P_k &=& \otimes_{k=1}^N \hat \sigma_k 
\eea
where $c_k$ are numerical constants, and $\hat \sigma_k$ are either Pauli spin operators $\hat x_k, \hat y_k, \hat z_k$
or the identity $\hat 1_k$. 
We can consider $\hat P_k$'s as elements of the UEA where the Lie algebra is a direct sum 
of $N$ $\LA{su}(2)$'s: $\mathcal{S} = \LA{su}(2)\oplus ... \oplus \LA{su}(2)$ and $\mathbb{K} = \mathbb{R}$. 
$\mathcal{S}$ is a semi-simple Lie algebra with $3N$ generators ($i\hat x_k, i\hat y_k, i\hat z_k$). 

There are $3^N$ CSAs for $\mathcal{S}$, which are based on selecting a particular 
Pauli operator ($\hat x, \hat y$ or $\hat z$) for each qubit and thus containing $N$ elements each, 
for example, $\{i\hat z_k\}_{k=1}^N$.

\subsubsection{Class-1 qubit mean-field Hamiltonians}

If one restricts unitaries to products of single-qubit operators 
\bea
\hat U_{\rm QMF} = \prod_{k=1}^N e^{i\tau_k(\bar{n}_k,\bar{\sigma}_k)},
\eea
where $\tau_k$ is an amplitude, $\bar{n}_k$ is a unit vector on the Bloch sphere, and $\bar{\sigma}_k = (\hat x_k,\hat y_k,\hat z_k)$.
The MF-solvable Hamiltonians within the qubit mean-field (QMF) approach\cite{Ryabinkin:2018di} are in the form of 
\eq{eq:MF1} with $\hat U_{\rm MF} = \hat U_{\rm QMF}$ and $\HC_k = i\hat z_k$ (or any other CSA of $\mathcal{S}$).
%where $\hat C_k$ are elements from one of the $\mathcal{S}$ CSA. 
One-qubit rotations in the qubit space do not translate to one-electron fermionic transformations (\eq{eq:fmf}) 
as was shown.\cite{Ryabinkin:2018di} Therefore, the QMF-solvable Hamiltonians are
 different from those in the $\LA{u}(N)$ algebra. 

Ladder operators of $\mathcal{S}$ are constructed as in the complex extension of the $\LA{su}(2)$ algebra: 
$\hat \sigma_{k}^{(\pm)} = \hat x_k\pm i \hat y_k$. 
$\hat U_{\rm QMF}$ transformation of CSA and ladder operators can be thought of as reorientation of individual 
quantization axes.   

\subsubsection{Qubit mean-field Hamiltonians of higher classes}

Due to distinguishability of qubits, higher classes of qubit MF-solvable Hamiltonians are somewhat simpler in their operator formulation 
than those of indistinguishable particles. These higher classes were reviewed extensively in the context of measurement 
problem in quantum computing.\cite{Izmaylov2019revising} It is easy to illustrate a higher class MF-solvable Hamiltonian starting 
with a simple two-qubit example: $\hat H_{\rm QMF,2} = \hat x_2 + \hat y_2 \hat z_1$. Here, all eigenstates can be taken as 
products $\ket{\phi_1\phi_2}$, where $\phi_i$'s are eigenstates for a single-qubit operator corresponding to the $i^{\rm th}$ qubit. 
One can easily confirm that the following 4 products are eigenstates of $\hat H_{\rm QMF,2}$ 
\bea
\{\ket{\uparrow_z \uparrow_{x+y}},~\ket{\uparrow_z \downarrow_{x+y}},~\ket{\downarrow_z \uparrow_{x-y}},~\ket{\downarrow_z \downarrow_{x-y}}\},  %~ \hat z \ket{\uparrow_z} = (+1)\ket{\uparrow_z},~
\eea
where $\ket{\uparrow_{q}}$ ($\ket{\downarrow_{q}}$) corresponds to the eigenstate of $\hat q$ with $+1(-1)$ eigenvalue. 
Note that depending on the eigenstate of the $\hat z_1$ operator the eigenstate for the second qubit changes.   

This consideration can be generalized to the $N$-qubit case, where
each qubit can have its quantization axis defined by a single qubit rotation and other qubits can have their own quantization 
axes that depend on whether previous qubit eigenstates are up or down along their quantization axes.  

Due to particle distinguishability, the criterion for the Hamiltonian to be MF-solvable can be formulated using the procedure of 
consequential integration of qubit variables with checking that there is always at least one single-particle operator commuting with 
each reduced Hamiltonian.\cite{Izmaylov2019revising} For $\hat H_{\rm QMF,2}$, there is $\hat z_1$ that commutes with $\hat H_{\rm QMF,2}$,
the reduced Hamiltonian is 
\bea
\hat H_r = \bra{\phi_1}\hat H_{\rm QMF,2} \ket{\phi_1} = \hat x_2 \pm \hat y_2,
\eea
where $\pm$ is determined by the choice of $\phi_1$, but in any case, there is a single qubit operator commuting with $\hat H_r$,
which is $\hat H_r$ because of its single-qubit nature.

\section{Conclusions} \label{conclusion}

%Fundamentally, our approach provides necessary and sufficient conditions for the Hamiltonian to 
%have all its eigen-states in the MF form, which is the most general form for un-entangled states.

We have proposed a general Lie-algebraic approach that provided necessary and sufficient conditions for 
defining the MF-solvable Hamiltonians. 
The generality of our consideration stems from its applicability to realization of the electronic Hamiltonian in any reductive Lie algebra.  
There are two main components of the algebraic consideration: 1) the largest commuting set of operators forming 
the Cartan sub-algebra and 2) the unitary transformations based on the universal covering Lie group obtained by 
exponentiating the Lie algebra used for the Hamiltonian realization.
Compared to previous considerations, our framework goes beyond simple quadratic fermionic Hamiltonians and also 
includes the Hamiltonians where multiple unitary MF rotations are needed to obtain all MF eigenstates.
 
The main practical value of the new definition is that it allows one to find decompositions of any Hamiltonian
into MF-solvable components. Such components has already been used in developing efficient quantum 
computing measurement scheme\cite{CSA2020} and will be generally useful in quantum simulations. Another advantage of the 
MF decomposition is in selecting the Lie algebra realization for the Hamiltonian of interest so that there is 
a predominant MF-solvable part. In this case doing perturbative expansions around the predominant MF-solvable 
part can provide an accurate treatment of many-body effects. Accuracy of this perturbative expansion 
will depend on magnitudes of residual couplings between neighbouring MF eigenstates 
compared to energy differences between these states.

\section*{Acknowledgements}
 A.F.I. is grateful to I.G. Ryabinkin, A.V. Zaitsevskii, and V.N. Staroverov for stimulating discussions and 
 acknowledges financial support from the Google Quantum Research Program, Zapata Computing, 
and the Natural Sciences and Engineering Research Council of Canada.

\section*{Appendix A: Compact Lie algebras from anti-hermitian operators}

Here we will show that any set of anti-hermitian operators that is closed with respect to the commutation 
operation, and whose elements have bounded spectrum, 
represents a compact Lie algebra. Also, such operators can be used to realize hermitian Hamiltonians. 
 
First, any set of anti-hermitian operators closed with respect to commutation form a real Lie algebra. 
If $\{\hat A_i\}$'s are basis elements of the Lie algebra of anti-hermitian operators, then 
\bea \label{eq:b}
[\hat A_i,\hat A_j] = \sum_k \xi_{ij}^{(k)} \hat A_k.
\eea
Let us obtain conjugate transposed version of this identity
\bea 
[\hat A_i,\hat A_j]^{\dagger} &=& \sum_k (\xi_{ij}^{(k)})^* \hat A_k^{\dagger} \\
-[\hat A_i,\hat A_j]&=& -\sum_k (\xi_{ij}^{(k)})^* \hat A_k \\ \label{eq:c}
[\hat A_i,\hat A_j]&=& \sum_k (\xi_{ij}^{(k)})^* \hat A_k.
\eea
Comparing Eqs.~\eqref{eq:b} and \eqref{eq:c}, 
it is clear that the structural constants must be real, $Im(\xi_{ij}^{(k)})=0$.

Second, to establish that a real Lie algebra is compact there should be a 
strictly positive definite bilinear form $(.,.)$ that satisfies the following condition
\bea 
([X,Y],Z) + (Y,[X,Z])=0,
\eea
where $X,Y,Z$ are elements of the Lie algebra. It was shown that the form $(X,Y) = Re{\rm Tr} (XY^\dagger)$,
satisfies these conditions for a set elements forming a real Lie algebra 
and closed under conjugate transpose.\cite{Knapp_comp} The only extra condition we need to introduce  
is that our operators have bounded spectra to avoid problems with the trace operation.

To use anti-hermitian operators for realizing the hermitian Hamiltonians $\hat H$ 
one can instead realize the anti-hermitian counterpart of the Hamiltonian $i\hat H$ and then multiply the 
obtained realization by $(-i)$. This extra factor will not change the validity of the maximal tori theorem 
applied to the compact part of the $\hat H$ realization.   

The Hamiltonians that can be expressed using only elements of the compact Lie algebra $\mathcal{A}$ 
(no quadratic and higher order terms in \eq{eq:H}) are MF-solvable. 
This statement relies on the maximal tori theorem for the compact Lie groups and algebras.\cite{Hall:MTT}  
The maximal tori theorem guarantees that for any compact Lie algebra $\mathcal{A}$, any of its elements can be transformed to 
CSA elements by a unitary transformation taken from the corresponding Lie group (\eq{eq:Us})
\bea\label{eq:MTT}
\hat U \left [ \sum_{k=1}^{|\mathcal{A}|} c_k \hat A_k \right ]\hat U^\dagger = 
\sum_{l=1}^{|\mathcal{C}|} b_l\HC_l, 
\eea
where $c_k$ and $b_l$ are all real (or purely imaginary) coefficients, and $|\mathcal{C}|$ is the number of CSA elements.

\section*{Appendix B: Partial and full MF-solvability}

We would like to illustrate that there are partially MF-solvable Hamiltonians. 
In other words, only a part of their eigenstates can be obtained using 
MF unitary transformations, and the rest of their spectra require going beyond the MF transformations. 
%The class of partially MF-solvable 
%Hamiltonians is similar to so-called quasi-exactly solvable problems if we add the constraint that only MF transformations can be used. 
Comments on the general procedure establishing full or partial MF-solvability of a Hamiltonian will follow after the example.

An example of a 2-electron Hamiltonian that is partially MF-solvable can be constructed using the condition that its eigenstates 
are product states if the first orbital is occupied
    \bea
        \hat H &=& \CR{1}\AN{1} + \hat H_{\text{non-MF}} (1-\CR{1}\AN{1})
    \eea 
    where a possible choice of the part that is not solvable by MF rotations is 
    \bea
        \hat H_{\text{non-MF}} &=& 
        0.43 \hat a_2^\dagger  \hat a_2 +
        0.15 \hat a_2^\dagger  \hat a_3 +
        0.81 \hat a_2^\dagger  \hat a_4 +
        0.64 \hat a_3^\dagger  \hat a_2 
        \nonumber \\ &-& 
        0.15 \hat a_3^\dagger  \hat a_2^\dagger  \hat a_3 \hat a_2 +
        0.54 \hat a_3^\dagger  \hat a_2^\dagger  \hat a_4 \hat a_2 -
        0.86 \hat a_3^\dagger  \hat a_2^\dagger  \hat a_4 \hat a_3 
        \nonumber \\ &+&
        0.89 \hat a_3^\dagger  \hat a_3 +
        0.21 \hat a_3^\dagger  \hat a_4 +
        0.66 \hat a_4^\dagger  \hat a_2 +
        \nonumber \\ &+&
                0.51 \hat a_4^\dagger  \hat a_2^\dagger  \hat a_3 \hat a_2 +
        0.76 \hat a_4^\dagger  \hat a_2^\dagger  \hat a_4 \hat a_2 +
        0.05 \hat a_4^\dagger  \hat a_2^\dagger  \hat a_4 \hat a_3 
       \nonumber \\ &+&   
        0.45 \hat a_4^\dagger  \hat a_3 +
        1.18 \hat a_4^\dagger  \hat a_3^\dagger  \hat a_3 \hat a_2 +
        0.25 \hat a_4^\dagger  \hat a_3^\dagger  \hat a_4 \hat a_2 
        \nonumber \\ &-&
        0.16 \hat a_4^\dagger  \hat a_3^\dagger  \hat a_4 \hat a_3 +
        0.68 \hat a_4^\dagger  \hat a_4.
    \eea 

A general procedure to establish Hamiltonian's MF-solvability would require a search for a MF unitary that provides at least one eigenstate of 
the given Hamiltonian. Minimizing variance of the Hamiltonian on state $\ket{0}$ containing some number of particles (i.e. physical vacuum) 
with respect to MF unitaries $\hat U$
\bea\label{eq:U1}
\hat U_1 = \arg\min_{\hat U} \bra{0} \hat U^\dagger \hat H^2 \hat U\ket{0} - \bra{0} \hat U^\dagger \hat H \hat U\ket{0}^2
\eea
allows one to determine whether at least one MF eigenstate can be generated by $\hat U_1$, 
which is the case if the variance can be lowered by $\hat U_1$ to zero.
If the variance cannot be lowered to zero, then the Hamiltonian is not MF solvable. Since the optimization is not linear, 
there can be complications related to finding a local minimum that does not correspond to zero variance even if zero variance exists 
as a global minimum within the space of all MF-unitaries. A random sampling of initial amplitudes of the MF rotations can be used 
to address this difficulty.     

If a zero-variance MF transformation is found, the next step is to determine whether this transformation provides all or only a subset of all eigenstates.
Applying the MF transformation to the Hamiltonian $\tilde H = \hat U_1^\dagger \hat H \hat U_1$ can answer this question: if $\tilde H = \tilde F_1(\hat C_k)$ contains only CSA elements then $\hat H$ is MF-solvable class 1, whereas if $\tilde H$ contains also a non-CSA part then further analysis is needed because $\hat U_1$ does not provide all eigenstates of $\hat H$. 

To go beyond $\hat U_1$ MF-eigenstates, let us assume that 
$\tilde H = \tilde F_1(\hat C_k) + \tilde F_2(\hat A_{k'})$, where $\tilde F_2$ is some polynomial over general Lie algebra 
elements (i.e. the non-CSA part). To determine whether there are other MF unitaries that can provide additional eigenstates 
one needs to factorize $\tilde F_1(\hat C_k) = F_1(\hat C_k) \hat P_1(\hat C_k)$ to projector $\hat P_1(\hat C_k)$ that defines the 
eigensubspace and an additional function $F_1(\hat C_k)$ (similar to \eq{eq:HMF2g}). 
The complimentary projector $\hat P_1^{\perp} = 1- \hat P_1$ can be identified and factorized from $\tilde F_2(\hat A_{k'}) = H_2(\hat A_{k'}) \hat P_1^{\perp}$. These factorization are always possible since $\tilde H$ has MF-eigenstates decoupled from the rest of the spectrum. 
To identify new MF rotations one needs to repeat the consideration that led to $\hat U_1$ starting 
from searching the minimum of the variance for the $ H_2(\hat A_{k'})$. This process as before can lead to three outcomes: 1) no MF rotations 
with zero variance are found, 2) zero-variance MF $\hat U_2$ is found, and it transforms $\hat H_2$ to a purely CSA polynomial, and 3) 
zero-variance MF $\hat U_2$ is found but it transforms $\hat H_2$ to a linear combination of a purely CSA and non-CSA polynomials. 
The first outcome means the end of the algorithm concluding only partial MF-solvability of $\hat H$. The algorithm can end in the second case as well, 
but with the conclusion that $\hat H$ is class-2 MF-solvable and its unitary rotations are $\hat U_1$ and $\hat U_1\hat U_2$. Finally, in the
third case the algorithm continues to search for other MF unitaries in the non-CSA part of the transformed $\hat H_2$.  There are two possible 
outcomes of this search: either $\hat H$ is partially MF-solvable or it is class $>$2 MF-solvable.

\section*{Appendix C: Non-trivial fermionic MF-solvable Hamiltonians}

Here, we present examples of class-1 and -2 MF-solvable Hamiltonians, for simplicity, they are limited to three spin-less orbitals. 
In both examples, CSA elements are taken as orbital occupation operators $\EE{p}{p}$. 

{\it Class 1:} Using form of \eq{eq:MF1} we set to zero all linear terms $c_k=0$, to have real coefficients in the Hamiltonian, $\phi_{pq} = 0$ 
in the MF rotation of \eq{eq:fmf}, while $\theta_{pq}$ and $d_{kk'}$ are chosen as follow
\bea\label{eq:d12}
  (d_{12}, d_{13}, d_{23}) &=& (-27, -9, -9) \\
  (\theta_{12}, \theta_{13}, \theta_{23}) &\approx& (-2.214, -1.459, -2.214) 
  %\\
  %U &=& \frac{1}{3} \begin{bmatrix}
  %    1 & -2 & 2 \\
  %    2 & -1 & -2 \\
  %    2 & 2 & 1
  %\end{bmatrix}.
  \eea
%All non-zero values of $\theta_{pq}$  and $d_{kk'}$ are round to $10^{-3}$.
 
This setup produces the following two-electron MF-solvable Hamiltonian  
\bea
  \hat H_{\text{MF}, 1} &=&
  11 \hat a_2^\dagger \hat a_1^\dagger \hat a_2 \hat a_1 
  +4 \hat a_2^\dagger \hat a_1^\dagger \hat a_3 \hat a_1 
  +4 \hat a_2^\dagger \hat a_1^\dagger \hat a_3 \hat a_2 
  \nonumber\\&&
  +4 \hat a_3^\dagger \hat a_1^\dagger \hat a_2 \hat a_1 
  +17 \hat a_3^\dagger \hat a_1^\dagger \hat a_3 \hat a_1 
  +8 \hat a_3^\dagger \hat a_1^\dagger \hat a_3 \hat a_2 
  \nonumber\\&&
  +4 \hat a_3^\dagger \hat a_2^\dagger \hat a_2 \hat a_1 
  +8 \hat a_3^\dagger \hat a_2^\dagger \hat a_3 \hat a_1 
  +17 \hat a_3^\dagger \hat a_2^\dagger \hat a_3 \hat a_2.\quad
\eea  

% here and below, the Hamiltonian coefficients are given with accuracy of $1\times10^{-3}$. 

    \textit{Class 2}: 
    For simplicity, we keep $\hat U_1 = I$ and $\hat P_1 = \hat E^1_1$ in Eq.~\cite{}.
    $\hat U_2$'s only non-zero rotation is $\theta_{23}=1.55$ so that $\hat U_2$ commutes with $\hat P_1$. 
    $F_1$ and $F_2$ are linear functions of CSA elements with coefficients \bea
        F_1 : \{c_k\} &=& \{0.44, 0.61, 0.95\} 
    \\
        F_2 : \{c_k\} &=& \{0.34, 0.69, 0.23\}.
    \eea These parameters generate the two-electron class-2 MF-solvable Hamiltonian whose all eigen-states are Slater determinants \bea
        \hat H_{\text{MF}, 2} &=& {
            -0.61 \hat a_2^\dagger \hat a_1^\dagger \hat a_2 \hat a_1
            -0.95 \hat a_3^\dagger \hat a_1^\dagger \hat a_3 \hat a_1
             +0.23 \hat a_2^\dagger \hat a_1^\dagger \hat a_2 \hat a_1 
        } 
        \nonumber \\ &+&  
            0.01 \hat a_2^\dagger \hat a_1^\dagger \hat a_3 \hat a_1 +
            0.23 \hat a_2^\dagger \hat a_2 +
            0.01 \hat a_2^\dagger \hat a_3 +
            0.01 \hat a_3^\dagger \hat a_1^\dagger \hat a_2 \hat a_1 
        \nonumber \\ &+&
            0.69 \hat a_3^\dagger \hat a_1^\dagger \hat a_3 \hat a_1 +
            0.01 \hat a_3^\dagger \hat a_2 +
            0.69 \hat a_3^\dagger \hat a_3
    \eea where the first two terms are $F_1(\hat C_k) \hat P_1$.

%\bibliographystyle{apsrev4-1}
%\bibliography{library}
%merlin.mbs apsrev4-1.bst 2010-07-25 4.21a (PWD, AO, DPC) hacked
%Control: key (0)
%Control: author (72) initials jnrlst
%Control: editor formatted (1) identically to author
%Control: production of article title (-1) disabled
%Control: page (0) single
%Control: year (1) truncated
%Control: production of eprint (0) enabled
%

\end{document}